\documentstyle[12pt]{article}

\if@twoside\oddsidemargin25mm
\else\oddsidemargin25mm\evensidemargin25mm\marginparwidth25mm\fi%
\newcommand{\dr}{\raise.3ex\hbox{$\stackrel{\leftarrow}{\delta }$}}
\newcommand{\dl}{\raise.3ex\hbox{$\stackrel{\rightarrow}{\delta}$}}
\newcommand{\beq}{\begin{equation}}
\newcommand{\eeq}{\end{equation}}
\newcommand{\eqn}[1]{(\ref{#1})}

\begin{document}

\begin{titlepage}
\begin{flushright} Preprint KUL-TF-95/14 \\
                   Preprint NIKHEF-95/025\\
                   hepth@xxx/9505174\\
                   May '95 \\
\end{flushright}
\vfill
\begin{center}
{\bf An alternative BRST operator for topological Landau-Ginzburg models }\\
\vskip 27.mm
{\bf F. De Jonghe$^{1}$}\\
\vskip   3mm
NIKHEF-H, Postbus  41882, 1009 DB Amsterdam, The Netherlands \\ \vskip 1cm
{\bf P. Termonia$^{2}$, W. Troost$^{3}$ and S. Vandoren$^{4}$ }\\
\vskip 3mm
Instituut voor Theoretische Fysica, K.U.Leuven
        \\Celestijnenlaan 200D, B-3001 Leuven, Belgium\\[0.3cm]
\end{center}
\vfill
\begin{center}
{\bf Abstract}
\end{center}
\begin{quote}
\small
We propose a new
BRST operator for the B-twist of $N=2$ Landau-Ginzburg
(LG) models. It solves the problem of the fractional ghost numbers
of Vafa's old BRST operator
and shows how the model is obtained by gauge fixing a zero action.
An essential role is played by the anti-BRST operator,
which is given by one of the supersymmetries of the $N=2$ algebra. Its
presence is needed in proving that the model is indeed a topological
field theory. The space of  physical observables,
defined by taking the anti-BRST cohomology in the BRST cohomology
groups, is unchanged.

\end{quote}
\vspace{2mm}
\vfill
\noindent $^1$ E-mail : t54@nikhef.nl \\
\noindent $^2$ E-mail: piet.termonia@fys.kuleuven.ac.be \\
\noindent $^3$ Onderzoeksleider N.F.W.O., Belgium\\
\hspace*{12pt}   E-mail: walter.troost@fys.kuleuven.ac.be\\
\noindent $^4$ E-mail: stefan.vandoren@fys.kuleuven.ac.be
\normalsize
\end{titlepage}
\section{Introduction}

Topological field theories (TFT) \cite{Witten,report} are
field theories with a BRST symmetry, and whose energy-momentum tensor is
BRST exact. Formally this implies, via the Ward identity, that the
partition function of the theory is independent of the metric on the
manifold on which the theory is defined. A large class of TFT's can be
constructed by gauge fixing a topological invariant
\cite{gftft} or by the so-called {\it twisting} of theories with
$N=2$ \cite{Witten,Twist} or $N=4$ \cite{Yamron} supersymmetry. This
twisting, in turn, can be done in two
different ways,  the so-called A- and B-twist \cite{ABTwist,BF}.
Let us consider the twisting of two dimensional $N=2$
Landau-Ginzburg models (LG).
These twists both involve changes in the  spins of the fermionic fields,
and the choice of a BRST operator, with the help of
the susy charges of one of the two the $N=2$ algebras \cite{Vafa}.
The relevant physical operators (observables) are representatives of the
BRST cohomology classes at some definite ghost number.
The assignment of these ghost numbers for the A-twist is
straightforward, but for the B-twist it is problematical. The most obvious
assignments lead to an action that has (in part) a ghost number different
from zero. A more elaborate assignment, involving ghost numbers that are
generically non-integer for most of the fields, allows for a consistent
ghost number zero action, but makes the subdivision of fields in classical
fields, ghosts and antighosts unclear (to say the least).

In this letter we intend to show that, by re-interpreting the customary
BRST charge for the B-twisted model as the sum of a BRST and an anti-BRST
charge, all ghost number assignments fall into place. We propose to take
for the BRST operator $one$ of the N=2 supersymmetry charges used by Vafa,
and the other as the anti-BRST operator. The corresponding
ghost number assignments
make a conventional separation in classical fields, ghosts
and antighosts straightforward, but the usual symmetry between BRST and
anti-BRST transformations is not present yet. The interpretation of the
anti-BRST transformation takes an entirely standard form, if one changes
to a different basis of fields, which is related in a (mildly) nonlocal
way with the customary basis.

As a
consequence of our procedure, the (++) component of the energy momentum
tensor is anti-BRST exact while the $(--)$ component is BRST exact. This
implies that
we also need the Ward identity for the anti-BRST operator in order to
prove that the theory is metric independent. Moreover, it also implies that
observables are subjected to two conditions, namely they should be BRST
invariant and their anti-BRST transformation should be BRST exact. This
leads us to define the physical spectrum as being the elements of the
anti-BRST cohomology defined in the BRST cohomology. We have computed the
spectrum in this way and it leads to the same topological observables
as in Vafa's approach.

\section{Anti-BRST and the topological twist}
We first remind the reader of the basic ingredients of the LG models, and
the topological models obtained from them by twisting, mainly to fix the
notation. Although a
description in an $N=2$ superfield formalism in two dimensions is
possible, the component notation seems most fit for the present purposes.
The Lagrangian of the model is
\begin{eqnarray}
&&S = \int d^2x\left[- \partial_+ X ^{i^*} \partial_- X^j\eta _{i^*j}
+2i \psi ^j \partial_- \psi ^{i^*}\eta _{i^*j}+2i \xi ^j \partial_+ \xi
^{i^*}\eta _{i^*j}\right.\\&&\left. \hspace{1cm}
+4\kappa \psi ^i \xi ^j \partial_i \partial_j W-4\kappa \psi ^{i^*} \xi
^{j^*} \partial_{i^*} \partial_{j^*} W^* - F^i F^{j^*} \eta _{ij^*}
-2 \kappa  F^{i^*} \partial_{i^*} W^* - 2 \kappa F^j \partial_j W
\right].\nonumber \label{LGact}
\end{eqnarray}
The bosonic fields $X$,$X^*$,$F$ and $F^*$ all are spin zero fields, while
the fermionic fields
$\psi$,$\psi^*$ have spin (helicity) $-{1 \over 2}$ and the fermionic fields
$\xi$,$\xi^*$
have spin ${1 \over 2}$. The potential $W(X)$ that determines the interaction,
is a quasi-homogeneous potential of degree $d $ with scaling weights
$\omega_i$, which means that for all complex $\lambda$,
\begin{equation}
       W \left( e^{\omega_i \lambda }X^i \right) = e^{d\lambda} W(X^i)    \,.
\end{equation}

This action possesses an $N=2$ global symmetry algebra. The
supersymmetry transformation rules are given by
\beq
\begin{array}{cc}
\delta X^i  =  \psi ^i \epsilon ^- + \xi ^i {\tilde \epsilon}^- &
\delta X^{i^*} = -\psi ^{i^*} \epsilon ^+ - \xi ^{i^*} {\tilde
\epsilon}^+  \\
\delta \psi ^i = - \frac{i}{2} \partial _+ X ^ i \epsilon ^+
- \frac{1}{2} F^i {\tilde \epsilon }^- &
\delta \psi ^{i^*}  =  \frac{i}{2} \partial _+ X^{i^*} \epsilon ^- -
  \frac{1}{2} F^{i^*}  {\tilde \epsilon }^+  \\
\delta \xi ^i  =  -\frac{i}{2} \partial _-X^i {\tilde \epsilon }^+
+\frac{1}{2} F^i \epsilon ^-  &
\delta \xi ^{i^*} = \frac{i}{2} \partial _-X^{i^*}{\tilde \epsilon }^-
+ \frac{1}{2} F^{i^*} \epsilon ^+  \\
\delta F^i = -i \partial_+ \xi^i  \epsilon ^+  + i \partial_- \psi^i
{\tilde \epsilon }^+ &  \delta F^{i^*}  = -i \partial_+ \xi^{i^*}
\epsilon ^-  + i \partial_- \psi^{i^*} {\tilde \epsilon }^- \, .
\label{SSTR}
\end{array}
\eeq
There are four different global parameters. When the supersymmetries
are made local,
the transformations close using the Lorentz transformations,
for which the $\psi$ and $\xi$ fields have spin $1\over2$ as mentioned.
Also there are two additional global $U(1)$ symmetries.
The charges of all the
fields are given in the table below (the index $i$ is suppressed but
understood in the table, and we abbreviated $h=\omega /d$.).
The symmetry for the $q_+$
charges is due to the quasihomogeneity of the potential.
\begin{table}[htf]
\label{tbl:charges} \begin{center}
\begin{tabular}{||c|c|c|c|c|c|c|c|c||}
\hline
     &$X$&$X^*$&$F$&$F^*$   &  $\psi $   &$\psi ^*$ &$\xi $ &$\xi ^*$ \\
     \hline
$q_+ $& $- 2h$ & $2h$&$2-2h$&$-2+2h$
&$1-2h$&$-1+2h $&$1-2h$&$-1+2h$ \\ \hline
$q_- $ & $0$ & $0$ & $0$ & $0$ & $1$ & $-1$ & $-1$ & $1$ \\ \hline
\end{tabular}\end{center}
\end{table}

In two dimensions, the Lorentz symmetry is a  $U(1)$ symmetry. The
basic idea behind the topological twist is that one can redefine the
Lorentz symmetry by combining the original one with the other
(global) $U(1)$ symmetries present in the model.  The two standard ways
\cite{ABTwist,BF} are
\begin{eqnarray}
\label{deftwist}
       s_A  =  s + {1 \over 2} q_+ && {\mbox   A-twist} \nonumber \\
       s_B  =  s  - {1 \over 2} q_-   && {\mbox   B-twist}  \, .
\end{eqnarray}
Here, $s$ denotes the spin of the field before the twist, i.e. for the
N=2 LG model, while $s_A$ and $s_B$ denote the spins in  the
A-twisted  and the B-twisted model respectively.
Note that the spins take on generically non-integer values for the
A-twist. In the B-twist (\ref{deftwist})
all fields have spin $s_B = 0$,  except the
fermions $\psi^i$, which have spin $s_B= -1$, and $\xi^i$, which have spin
$s_B= 1$. We will focus on the B-twist from now on.

This redefinition of spin has two main consequences.
The first is that the coupling to gravity changes,
i.e. there is a change in the energy-momentum tensor. The second is
that two of the four supersymmetries now have zero $s_B$ spin,
namely those
parametrised by $\epsilon^+$ and by $\tilde \epsilon^+$.
Following \cite{Vafa}, these two supersymmetries can be used to
construct a spinless fermionic operator
$\delta$ (acting from the left):
\begin{eqnarray}
\delta X^{i^*}=\psi ^{i^*}+\xi ^{i^*} &\qquad& \delta F^i=i(\partial _+\xi
^i-\partial _-\psi ^i)\nonumber\\
\delta \xi ^{i^*}=\frac{1}{2}F^{i^*} &\qquad& \delta \xi
^i=-\frac{i}{2}\partial _-X^i\nonumber\\
\delta \psi ^{i^*}=-\frac{1}{2}F^{i^*} &\qquad& \delta X^i=0\nonumber\\
\delta F^{i^*}=0 &\qquad& \delta \psi ^i=-\frac{i}{2}
\partial _+X^i \ . \label{BRST}
\end{eqnarray}
{}From these expressions it is obvious that $\delta^2=0$. It is proposed in
\cite{Vafa} to interpret $\delta $ as a BRST operator of a so far
unspecified gauge symmetry. The action of the LG model
can be written as
\begin{eqnarray}
S&=&4\kappa \psi ^i\xi ^j\partial _i\partial _jW-2\kappa F^i\partial
_iW\nonumber\\
&&+\delta [F^i(\psi ^{i^*}-\xi ^{i^*})+4\kappa \partial _{i^*}W^*\psi
^{i^*}+iX^{i^*}(\partial _-\psi ^i+\partial _+\xi ^i)]\nonumber\\
&\equiv &S^0+\delta \Psi \ .\label{TLG}
\end{eqnarray}
This is of the same form as a classical action, supplemented by a gauge
fixing action which is the BRST variation of a gauge fermion. It is still
not specified what the gauge symmetry of this action would be.
One can show that the energy-momentum tensor is BRST exact, and therefore
equivalent to zero, which makes the theory into a metric-independent one,
i.e. a topological field theory.

If one tries to supply the missing ingredients to make contact with
the usual gauge
fixing approach to TFT, where the operator $\delta$ clearly is to be
interpreted as a BRST operator of  a closed symmetry algebra, one
encounters a problem. This can most easily be seen by trying to figure
out the ghost numbers of all the fields. They have to be chosen such
that the action has ghost number
zero and the BRST operator raises ghost number by one.
There is one solution to these requirements, namely
to assign  ghostnumbers equal to minus the $q_+$-charge
\cite{PD}.
These assignments are unsatisfactory however. The fact that all
are non-integer is uncommon, although not necessarily wrong.
One could imagine that, due to an anomaly in the ghost number current
of the quantum theory, shifts of the ghost number are induced.
We are working on the classical level however, where such a feature seems
impossible in the conventional approach. There the ghost number assignments
follow from a definite procedure, and necessarily come out to be
integer (positive or negative). Thus an interpretation as a gauge fixed
theory is impossible.
If one tries to rescue this, by assigning integer ghost numbers to all
the fields, such that the BRST operator has ghost number one, then the
action contains terms of  ghost number minus 2 \cite{BF}.

Another approach is to try and find the gauge symmetry directly.
Looking at $S^0$ in \eqn{TLG}
one  of the gauge symmetries is a certainly  a shift in $X^{i*}$, and one
would therefore introduce a ghost field for this symmetry.
Looking at the transformation rule eq.(\ref{BRST})
for $X^{i*}$, one  sees that in fact we have introduced
{\em two} ghost fields for only one symmetry. The theory is then
reducible, and $F^{i*}$ are the ghosts for ghosts.
This seems to be an unnecessary
complication.
The BRST algebra for this
symmetry is the left column of \eqn{BRST}.
For the right column, the interpretation is not so clear. There seems not
to be a gauge symmetry and a corresponding action, for which the right column
is the BRST algebra.

To remedy this situation, {\it we propose to change the BRST operator}.
The previous BRST operator was obtained from the supersymmetries
with as BRST parameter $\Lambda=\epsilon^+ = \tilde \epsilon^+$.
Instead, we propose to use simply the first of these supersymmetries, and
interpret it as a BRST operator by itself. The second supersymmetry we
propose to identify with the anti-BRST operator%
\footnote{For a review of
the use of BRST--anti-BRST symmetry of gauge theories, we refer to
\cite{BAB}.}%
. We will call these
operators $\bf s$ and $\bf{\bar s}$ respectively.
The transformation rules are:
\begin{equation}
  \begin{array}{ll}
   \bar {\bf s} X^{i*} =  \psi^{i*}  & {\bf s} X^{i*} = \xi^{i*} \\
   \bar {\bf s} \psi^i =  -\frac{i}{2} \partial _+ X^i & {\bf s} \psi^{i*}
   = -\frac{1}{2} F^{i*} \\
   \bar {\bf s} \xi^* = \frac{1}{2} F^{i*}  & {\bf s} \xi^i =  -\frac{i}{2}
\partial _- X^i \\
   \bar {\bf s} F^i = i \partial_+ \xi^i & {\bf s} F^i = -i \partial _-
   \psi^i \label{ouranti}\ ,
 \end{array}
\end{equation}
with all the other (anti)BRST transformations vanishing.
One easily
verifies the important nilpotency relations ${\bf s}^2 = {\bf {\bar s}}^2 =
{\bf s} {\bf {\bar
s}} + {\bf {\bar s} s} = 0$. Comparing with \eqn{BRST} we see that
the BRST operator introduced by Vafa is the sum,
$\delta = \bf s + \bf{\bar s}$.
The invariance of the action under $s$ and $\bar s$ follows of course from
the original supersymmetries.

The condition that fixes the ghost number assignments
is now that $s$ raises the ghost number by one unit,  $\bar s$
lowers it by one unit, and the action has ghost number zero.
All ghost numbers are integers.
In fact, the ghost number turns out to be nothing but the $q_-$ charge
(see table).

With this new interpretation,
the  action of the LG model can still be   written
as the sum of  a classical action and a gauge fixing part.
One easily computes
\begin{eqnarray}
  S & = &  4\kappa \psi ^i\xi ^j\partial _i\partial _jW-2\kappa F^i\partial
      _iW \nonumber \\
    && + {\bf s \bar s } \left[4 \kappa W^* +2 X^{i*} F^i \right] \ .
    \label{TLGA}
\end{eqnarray}
The classical part does not depend on $X^{i*}$,
and therefore one has a gauge (shift-)symmetry
$\delta X^{i*} = \varepsilon^{i*}$,
and the corresponding ghosts $\xi^{i*}$. In accordance with the spirit of
the BRST--anti-BRST scheme \cite{BAB}, one introduces also an
antighost $\psi^{i*}$, and its BRST variation $F^{i*}$.
Apart from this quartet, there is a second set of fields transforming
into each other, viz.  $F^i, \psi^i, \xi^i$ and $X^i$.
The reason for the presence of the latter fields, and for their
transformations,
eq.(\ref{ouranti}), is not obvious at this stage, but we will come back to
their interpretation.
It is now clear that the gauge fixing part ${\bf \bar s s} K$
fixes the  shift symmetries, as one would do starting from a
zero action to construct a TFT \cite{gftft}.

The identifications above do not yet exhibit the  usual structure of
BRST-anti-BRST.
A first signal is that the first term in \eqn{TLGA} should not be present
in the underlying gauge invariant classical action, since $\psi ^i$ and
$\xi ^i$ have non zero ghost number, and the classical action supposedly
depends only on classical fields. This term should rather be a part of
a gauge fixing term instead. A second point is that there should be
more symmetry between ghosts and antighosts. The anti-BRST transformation
of the classical fields are identical to their BRST transformation,
when replacing ghosts with antighosts. This is not the case for the
second set, since we then also have to interchange $\partial _+$ and
$\partial _-$.

In the $N=2$ LG model,  the starred and unstarred fields occur
symmetrically.
The twist has lifted this symmetry: the former are all spinless, but
$\psi ^i$  and $\xi ^i$ have helicities 1 and -1 respectively.
One can construct $\psi ^idx^+$ and $\xi ^idx^-$, which behave as one
forms under holomorphic coordinate transformations. The asymmetry is
mirrored in
the derivatives in the transformation laws for the second set, which is in
accordance with the helicity-assignment. At the same time, one can also
consider
$F^i$ to be a two-form, which can not be distinguished from a scalar
in the treatment with a flat metric.
The BRST--anti-BRST symmetry can be redressed by the
following non-local change of field variables:
\begin{eqnarray}
\psi ^i&=&\partial _+\chi ^i\nonumber\\
\xi ^i&=&\partial _-\rho ^i\nonumber\\
F^i&=&\partial _-\partial _+H^i  \ .
\end{eqnarray}
All the fields on the right hand side are scalars.
Remark that the Jacobian of this transformation is one, at least formally,
since the
contributions from the fermions cancel against the bosons.
For the new variables we can take the transformation rules
\begin{eqnarray}
\bar {\bf s} \chi^i= -\frac{i}{2}X^i &
\qquad& {\bf s} \rho^i =
-\frac{i}{2}X^i \nonumber\\
\bar {\bf s} H^i = i \rho^i &\qquad& {\bf s} H^i = -i\chi^i \ ,
\label{qaq for nonlocal}
\end{eqnarray}
to reproduce the so far unexplained rules in (\ref{ouranti}).
They now correspond to a shift symmetry for the field $H^i$, introducing
the ghost field $\chi^i$. The antighost is $\rho^i$, and $X^i$
completes the quartet.
It is clear that we  have uncovered a manifest BRST anti-BRST symmetry.

The action, when written in terms of the new fields, is BRST exact:
\begin{eqnarray}
  S&=&{\bf s}[4\kappa i\partial _+H^i\partial _-\rho ^j\partial _i\partial
_jW]+{\bf s \bar s } \left[4 \kappa W^* +2 X^{i*} F^i \right]\nonumber\\
&=&{\bf s}[\xi ^i(-2i\partial _+X^{i*}+4\kappa i\partial
_+H^j\partial _i\partial _jW)+\psi ^{i*}(2F^i+4\kappa \partial _{i*}W^*)]\
. \label{S is exact}\end{eqnarray}
This allows the following interpretation. One starts from two classical
fields, $X^{i^*}$ and $H^{i}$.
The classical action is zero, and the symmetries are shift symmetries,
with ghosts $\xi ^{i*}$ and $\chi ^i$.
Then one introduces  antighosts  $\psi ^{i^*}$ and $\rho ^i$, and
Lagrange multipliers  $X^i$ and $F^{i^*}$. This completes the field
content of the theory. The gauge fixed action is the  BRST exact
functional, given in eq.(\ref{S is exact}).
Note that the actual content of the resulting TFT depends heavily on
the gauge fixing procedure, as usual: there are no physical local
fluctuations, but global variables may remain. We have nothing to add on
this point, so we refer to the existing literature\cite{Vafa}.

We conclude that our proposal for the BRST operator has led,
via the transformation of \eqn{qaq for nonlocal},
to a re-interpretation of the
B-twisted Landau-Ginzburg model which makes contact with the
alternative view on topological field theories, as arising from gauge
fixing a zero action.

Having changed the BRST operator, we now discuss the implications of this
change. First of all, we investigate wether we still have a topological
theory in the sense that the energy-momentum is BRST exact for the new BRST
operator. Afterwards, we will investigate whether the physical content
(observables) of the theory has changed.

\section{The energy-momentum tensor}

To prove the topological nature of the theory we have to show that
the energy momentum tensor is trivial.
To compute it, it is not necessary
to covariantize the (flat space) LG action \eqn{LGact} completely.
The transformation properties of the various fields can be seen by
considering {\em holomorphic} changes of coordinates only.
One finds that $\psi ^i dx^+$  and $\xi ^i dx^-$ are $(1,0)$ and $(0,1)$
forms respectively, which we assemble in
$A^i= \psi ^i dx^+ +\xi ^i dx^-$ purely for notational convenience.
Likewise, $F^i$ is a $(1,1)$ form component (we use the bold symbol for
the form itself). The BRST and anti-BRST transformation respect the form
degrees. The covariant action is
\begin{eqnarray}
S&=&\int \frac{1}{2}dX^i\wedge[dX^{i*}]_*+iA^i\wedge d(\psi ^{i*}-\xi
^{i*})-iA^i\wedge [ d(\psi ^{i*}+\xi ^{i*})]_*\nonumber\\
&&+\int2\kappa A^i\wedge A^j\partial _i\partial _jW-4\kappa \psi
^{i*}\xi ^{j*}\partial _{i*}\partial _{j*}W^*\sqrt gdx^+\wedge
dx^-\nonumber\\
&&- \int F^{i*}{\bf F}^i+2\kappa F^{i*}\partial _{i*}W^*\sqrt gdx^+\wedge
dx^-+2\kappa {\bf F}^i\partial _iW\ ,
\end{eqnarray}
where the lower star is used to denote the Hodge dual,
$[\omega ]_*=\sqrt gg^{\mu \nu }\epsilon
_{\nu \rho }\omega _\mu dx^\rho $.
The metric dependence is in the volume element
and in the definition of the Hodge dual. The derivative of this
action w.r.t. the metric gives the energy momentum tensor~:
\begin{eqnarray}
     T^B_{++}=&- \partial_+ X ^{i^*} \partial_+ X^i + 2i \psi^i
     \partial_+\psi^{i*}&=\bar {\bf s} \left[ 2i \psi^i \partial_+ X ^{i^*}
     \right] \nonumber \\
     T^B_{--}=&- \partial_- X ^{i^*} \partial_- X^i + 2i \xi^i
     \partial_- \xi^{i*}&={\bf s} \left[ 2i \xi^i \partial_- X ^{i^*}
     \right] \\
     T^B_{+-}=&4\kappa \psi ^{i^*} \xi ^{j^*}
     \partial_{i^*} \partial_{j^*} W^* +2 \kappa  F^{i^*} \partial_{i^*}
     W^*&={\bf s \bar s} \left[4\kappa W^* \right] \nonumber \ .
\end{eqnarray}
After the derivation, we have taken the metric to be flat. These are
therefore the relevant operators for variations of correlation functions
around a flat metric.

The (++) component is  anti-BRST exact, but not BRST exact in spite of
\eqn{S is exact}.
The reason is
that the BRST operator depends on the metric and one cannot commute the
BRST variation and the derivative w.r.t. the metric \cite{tftbv}.

To prove metric independence of correlation functions, one needs
not only BRST invariance, but also
the Ward identity for the anti-BRST operator.
What is needed is that the physical operators are
BRST invariant, {\em  and} that their anti-BRST variation is BRST
exact. In that case one can argue as follows.
Denoting by ${\cal O}_i,
i=1,...,N$ a set of (metric independent) physical operators that satisfy
${\bf s}{\cal O}_i=0, {\bar {\bf s}}{\cal O}_i={\bf s}V_i$
and $T_{++}={\bar {\bf s}}X_{++}$~:
\begin{eqnarray}
\frac{\delta }{\delta g^{++}}<{\cal O}_1...{\cal O}_N>&=&<T_{++}{\cal
O}_1...{\cal O}_N>\nonumber\\
&=&<{\bar {\bf s}}X_{++}{\cal O}_1...{\cal O}_N>\nonumber\\
&=&<X_{++}\sum _{i=1}^N{\cal O}_1...{\bf s}V_i...{\cal O}_N>\nonumber\\
&=&<({\bf s}X_{++})\sum _{i=1}^N{\cal O}_1...V_i...{\cal O}_N>\nonumber\\
&=&<\psi ^iy_{\xi ^i}\sum _{i=1}^N{\cal O}_1...V_i...{\cal O}_N>\ ,
\end{eqnarray}
where $y_{\xi ^i}\equiv \frac{\dr S}{\delta  \xi ^i}$.
We have assumed that there are no BRST nor anti-BRST anomalies.
In order to have a topological field theory the result
should vanish.
Classically, one may use the field equations $y=0$.
In fact, also in the formulation of \cite{Vafa}, field equations were
used implicitly. The (++) component of his energy
momentum tensor is only BRST exact modulo the {\it same} field equations,
i.e. $T_{++}=\delta (-2i\partial _+X^{i*}\psi ^i)+\psi ^iy_{\xi ^i}$.
On the other hand, in the quantum theory, these field equations might
contribute to order $\hbar$. This contribution
can in principle be computed using the Schwinger--Dyson equations.
This formally leads to
products of operators at the same space-time point.
To make a proper computation
one needs a regularisation scheme.
That computation
is beyond the scope of this paper, but is currently under study
\cite{RegTwist}.

The proper cohomological formulation is, that one first determines the
${\bf s}$ cohomology, a space of equivalence classes.
The operator ${\bar {\bf s}}$ is well defined and
nilpotent in that space, so that {\em its} cohomology can be used
as our characterisation of physical states%
\footnote{In the usual gauge theories, the more
common procedure of choosing an anti-BRST invariant representative in
each equivalence class amounts to the same thing.}.
This characterisation is not arbitrary, but more or less forced upon us
by the requirement that the energy momentum tensor is trivial. We now
investigate this cohomology.

\section{The spectrum}

The observables in  Vafa's
picture are  the solutions of the $\delta $-cohomology, while
in our interpretation they are the solutions of the ${\bar {\bf
s}}$-cohomology in the ${\bf s}$-cohomology.
Let us for example consider  observables which are integrals
of functions
$\int \Phi^{(2)}$, over the Riemann surface. For simplicity we restrict
ourselves
here to integrals over $(1,1)$-forms with respect to holomorphic coordinate
transformations, although the two spectra  coincide even if one imposes
no restriction at all. For the $\delta $ operator, we have to solve the
descent equations~: \begin{eqnarray}
\delta \Phi ^{(2)}&=&-{\bf d}\Phi ^{(1)} \ ,\nonumber\\\
\delta \Phi ^{(1)}&=&-{\bf d}\Phi ^{(0)}\ ,\nonumber\\
\delta \Phi ^{(0)}&=&0\ ,
\label{desdel}
\end{eqnarray}
where ${\bf d}$ is the graded exterior derivative, defined as
$(-1)^{gh}$d, and $\Phi ^{(k)}$ is a $k$-form.
The equalities are always taken modulo field equations, i.e. we compute
the weak cohomology.
We denote by $M$ the space of  formal sums
$\Phi = \Phi^{(2)} + \Phi^{(1)} + \Phi^{(0)}$.
The descent equations take the form:
\begin{equation}
(\delta + {\bf d}) \Phi = 0.
\end{equation}
and the relevant cohomology is translated into the
$\delta + {\bf d}$ cohomology.
The solution is given by the polynomial ring of the LG potential
\cite{BF}~: \begin{eqnarray}
\Phi^{(0)} &=& P(X)\ ,\nonumber\\
\Phi ^{(1)}&=&-2i\partial _iP(\psi ^idx^++\xi ^idx^-)\
,\nonumber\\
\Phi ^{(2)}&=&[-4\partial _i\partial _jP\psi ^i\xi ^j+2\partial
_iPF^i]dx^+\wedge dx^- \ ,\label{delobs} \end{eqnarray}
where $P(X)$ is a polynomial corresponding to some non trivial element of
the ring determined by the potential $W(X)$. Indeed, the vanishing
relations $\partial _iW=0$ follow from the field equation of $F^i$, which
imply that $\kappa \partial _iW $ is weakly equal to $ \delta \psi ^{i*}$.

Now we turn to the ${\bf s}$ cohomology.
In a first step,
$H({\bf s},M)=\frac{ker({\bf s}+{\bf d})}{im({\bf s}+{\bf d})}$. Since
${\bf d}$ is the graded exterior derivative, $\bar {\bf s}+{\bf
d}$ is well defined on $H({\bf s},M)$,
and we can calculate
its cohomology in $H({\bf s},M)$.
The physical observables are then given by
$H(\bar {\bf s},H({\bf s},M))= \frac{ker(\bar {\bf s} + {\bf d})}%
{im(\bar {\bf s} + {\bf d})}$. The classes
of $H(\bar {\bf s},H({\bf s},M))$
are represented by those forms
$\Phi$ for which there exists
a $Y = Y^{(2)} + Y^{(1)} + Y^{(0)}$ such that
\begin{eqnarray}
({\bf s}+{\bf d})\Phi=0 && (\bar {\bf s} + {\bf d}) \Phi = ({\bf s} + {\bf
d} ) Y \ ,
\end{eqnarray}
and which itself is not ${\bar {\bf s} + {\bf d}}$-trivial. The solution is
given by:
\begin{eqnarray}
\Phi^{(2)}=(-4\partial_i\partial_jP \psi^i \xi^j + 2
\partial_{i} P F^i) dx^- \wedge dx^+ && Y^{(2)}=-\Phi^{(2)}\ , \nonumber \\
\Phi^{(1)}=2i \partial_i P \psi^i dx^+ && Y^{(1)}=-2i
\partial_i P \xi^i dx^-\ , \nonumber\\
\Phi^{(0)}=0 && Y^{(0)}=P\ .
\end{eqnarray}
This shows that the spectra coincide. Needless to say, the same result
is true if we interchange the order in which the cohomologies are computed.

\section{Conclusions}

The main observation of the present paper is  that in
the topologically (B-)twisted  2d Landau-Ginzburg model, the
topological symmetry is most naturally interpreted in terms of a
BRST--anti-BRST symmetry.
This  interpretation  allows one to reconcile the requirements of integer
ghost number, and ghost number zero for the action.

The anti-BRST symmetry has been used in TFT to study
topological Yang-Mills theory \cite{antifix}, and also recently in
a more general context
\cite{Baulieu}. In these cases, it seems more a matter of choice whether
one mentions the anti-BRST symmetry or not.
In contrast, for our proposal the anti-BRST
is forced upon us in order to make sense of the twisted model as
the gauge fixing of a zero action, and also from the requirement that
the model is topological, so that its energy momentum tensor must be
``trivial''.
Finally
we want to remark that this procedure can  be
extended to the interpretation of the B-twist of any
model 2d $N=2$ model. Indeed, the interpretation of the $N=2$ SUSY
transformations in terms of (anti)BRST transformations is much more
general then the precise model that we used, since it only relies on the
off-shell formulation of the
$N=2$  algebra  (using auxiliary fields). This formulation of the
algebra is the same in e.g. $\sigma$-models, such that there too a more
natural interpretation of the twisted model may be based on the
BRST--anti-BRST symmetry.

\vspace{0.5cm}

ACKNOWLEDGEMENT:\\
We gratefully acknowledge interesting discussions with P.
Fr\`e.

This work was carried out in the framework of the project "Gauge theories,
applied supersymmetry and quantum gravity", contract SC1-CT92-0789 of the
European Economic Community. The work of FDJ was supported by the Human
Capital and
Mobility Programme by the network on {\it Constrained Dynamical Systems}.


\end{document}